\documentclass[twoside]{dis08}
\usepackage[latin1]{inputenc}
\usepackage[dvips]{graphicx,epsfig,color}
\usepackage{wrapfig,rotating}
\usepackage{amssymb,amsmath,array}

\begin{document}
\title{Transverse Spin Physics at COMPASS}

\author{Federica Sozzi$^1$ on behalf of the COMPASS Collaboration
%
%
\vspace{.3cm}\\
%
1- Trieste University and INFN, \\
via Valerio 2 34127 Trieste  Italy
}

\maketitle

\begin{abstract}
The study of transverse spin effects
is part of the scientific program of COMPASS,
 a fixed target experiment at the CERN SPS.
COMPASS investigates the transversity PDFs in semi-inclusive DIS,
using a  longitudinally polarized muon beam  of 160~GeV/c
impinging on a transversely polarized target.
From 2002 to 2004, data have been collected using a  $^6$LiD target transversely polarized.
Transversity has been measured using different
quark polarimeters: the azimuthal distribution of single hadrons,
the azimuthal dependence of the plane containing hadron pairs,
and the measurement of the transverse polarization of baryons
($\Lambda$ hyperons).
All the asymmetries have been found to be small,
and compatible with zero,
a result which has been interpreted as a cancellation
between the u and d-quark contributions.
In 2007 COMPASS has taken data using a NH$_3$ polarized 
proton target which will give complementary information on  transverse spin effects.

\end{abstract}

\section{Introduction}

To fully specify the quark structure of the nucleon at twist two level, 
three parton distribution functions are needed: the well known momentum 
   and  helicity distribution $q(x)$ and $\Delta q(x)$, 
and the transversity distribution $\Delta_T q(x)$.
This latter describes the probability density of finding transversely polarized
quarks in a transversely polarized nucleon, 
and it is currently receiving a lot of attention, both from the experimental and from 
the theoretical point of view. 

Due to its chiral odd nature, the transversity PDFs must be coupled to another chiral-odd 
function in order to build an observable.
For this reason, 
  transversity cannot be measured in   DIS experiments, 
but only in semi-inclusive DIS experiments (SIDIS), 
where at least one hadron in the 
final state is detected.
In SIDIS, different channels 
allow to access transversity: the azimuthal distribution of single hadrons, 
the azimuthal dependence of hadron pairs and the measurements 
of the transverse polarization of 
$\Lambda$ hyperons. All these channels have been investigated by the COMPASS 
Collaboration.

COMPASS~\cite{Abbon:2007pq} is a fixed target experiment at the CERN SPS, 
with a physics program focused on the study of the nucleon spin structure.
From 2002 to 2004, COMPASS took data with a longitudinally polarized 
muon beam of 160~GeV/c and a $^6$LiD target. 
For about the 20\% of the time, the target nucleons have been polarized transversely 
with respect to   the beam direction, in order to allow the measurements of transverse spin 
effects. 
The $^6$LiD material is characterized by a high dilution factor ($f\sim$0.38), 
and the polarization values  P$_t$ reached during data taking are around 50\%.

\section{Single spin asymmetries}
\subsection{Collins asymmetries}
 The transversity PDF can be accessed in single hadron production 
via the so called Collins effect~\cite{Collins:1993kk}.
In this mechanism, the fragmentation 
 function of transversely polarized quarks into unpolarized hadrons 
is composed  by two parts, 
an unpolarized  and a polarized one, depending on the quark spin (Collins FF).   
More precisely, the FF shows an azimuthal dependence 
with respect to the plane defined by the quark momentum and the quark spin, 
meaning that the event yield can be written as:
\begin{equation}
N = N_0 \cdot (1+f \cdot P_t \cdot D_{nn} \cdot A_C \cdot \sin(\phi_C)) 
\label{eq:collfun}
\end{equation}
where $f$ and  P$_t$ have been already introduced, and $ D_{nn}  = (1-y)/(1-y+y^2/2)$ is the transverse spin
transfer coefficient from the initial to 
the struck quark. 
The Collins angle $\phi_C$ is defined as $\phi_h - \phi_{s'}$, where $\phi_h$ is the angle of the transverse momentum of the outgoing hadron and $\phi_{s'} = \pi - \phi_{s}  $ is the azimuthal angle of the struck quark spin ($\phi_{s}  $ is the azimuthal angle of quark  before the hard scattering). 
$A_C$ is the Collins asymmetry, proportional to the convolution of the 
  Collins fragmentation function and the transversity distribution:
\begin{equation}
A_C= \frac{\sum_q e_q^2 \, \Delta_T q(x) \, \Delta_T^0 D_q^h(z, p_T^2)}
       {\sum_q e_q^2 \,     q(x) \,       D_q^h(z, p_T^2)} 
\end{equation}
where  $z= E_h /(E_{l}-E_{l'})$ is the fraction of available energy carried by the hadron,
and $p_T$ is the hadron transverse momentum with respect to the virtual photon direction.
Comparing the number of hadrons produced in SIDIS reactions on    
nucleons polarized transversely in opposite directions, 
the modulation given by the Collins angle gives access to  the asymmetry $A_C$.

In order to select DIS events in the data, 
 interaction vertices with one incoming muon, one scattered muon and at least one outgoing hadron
have been selected. To select the DIS regime, the photon virtuality
$Q^2$  is taken above 1 (GeV/c)$^2$, 
the fractional energy of the virtual photon $y$   between 0.1 and 0.9, and 
the invariant mass of the final hadronic state $W$ above 5 GeV/c$^2$.
Moreover, the following cuts on the  hadron variables have been applied: $z>0.2$ 
and $p_T >$ 0.1 GeV/c.
After all the cuts applied, the total number of positive and negative hadrons is
7 and 8.5$\cdot 10^6$, respectively. 

\begin{wrapfigure}{r}{0.55\columnwidth}
\centerline{\includegraphics[width=0.55\columnwidth]{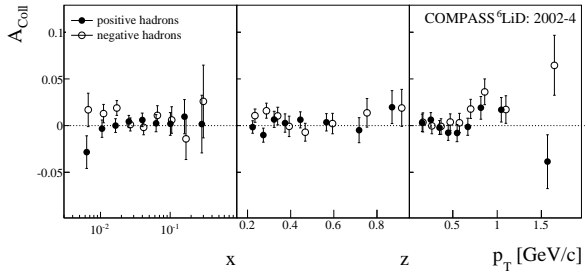}}
\caption{Collins asymmetries for not identified hadrons as a function of $x$, $z$ and $p_T$.}\label{Fig:nonid}
\end{wrapfigure}

The Collins  asymmetries of unidentified hadrons~\cite{Alexakhin:2005iw, Ageev:2006da} 
for the full statistics are shown  in Fig.~\ref{Fig:nonid}, as a function of 
 $x$, $z$ and $p_T$. 
The asymmetries are small, and
compatible with zero within the very good statistical accuracy, 
which is reaching $\sim$1\% in the central $x$ bins. 
The systematic errors, investigated with extensive studies,
have been found to be well below the statistical precision.
 A na{\"\i}ve interpretation of the results can be achieved in the parton model: 
restricting to the valence region,
 the small asymmetries on deuterium  are due to a cancellation between  
the transversity PDF for the u and d quark. This result is valid even if the 
Collins FF are opposite in sign and similar in size, as is suggested by the 
non zero   asymmetries on a proton target measured by the HERMES Collaboration~\cite{Diefenthaler:2007rj}.
\begin{wrapfigure}{r}{0.55\columnwidth}
\centerline{\includegraphics[width=0.55\columnwidth]{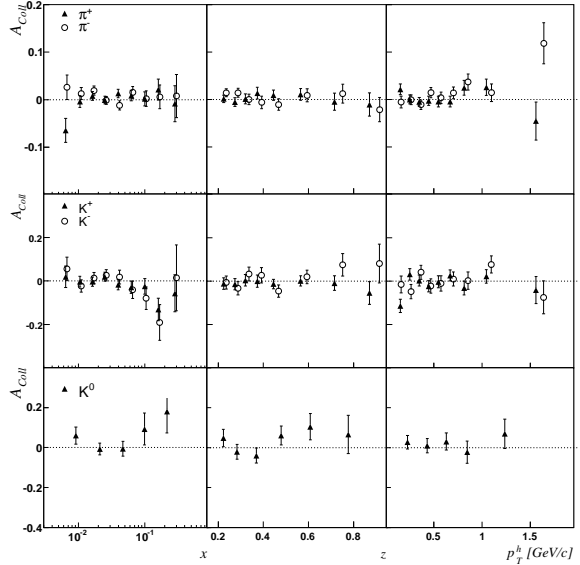}}
\caption{Collins asymmetries for pions and kaons as a function of $x$, $z$ and $p_T$.}\label{Fig:pik}
\end{wrapfigure}

 In order to perform a more sophisticated flavor separation analysis, 
  the hadrons in the final state have to be identified. 
  In COMPASS, charged  kaons and pions are identified a  RICH~\cite{Albrecht:05}.
 Due to the different Cherenkov thresholds, 
pions and kaons can be identified for momenta larger than $\sim$3~GeV/c and 10~GeV/c, respectively,
and up to 50~GeV/c, that corresponds to a 1.5~$\sigma$ $\pi$-K mass separation. 
Neutral kaons  have been selected in the data from the invariant mass of the decay products. 
After the selection, the invariant mass spectra show a peak at the $K^0$ mass,
with a signal-to-background  ratio around 15.
The final sample of $K^0$s for the analysis has been selected  considering a $\pm$20~MeV/c$^2$ 
mass window around the $K^0$ mass.
The final statistics consists of 5.2 and 4.5$\cdot 10^6$  $\pi^+$ and $\pi^-$, 0.9 and 0.6$\cdot 10^6$ $K^+$ and $K^-$, 
and 0.25$\cdot 10^6$  $K^0$.
The results obtained for pions and kaons~\cite{:2008dn} are presented
 in Fig.~\ref{Fig:pik}.
Also here, 
the asymmetries are small and compatible with 
zero within the statistical errors. 
 
\subsection{Sivers asymmetries }
The Collins asymmetry is not the only possible single spin asymmetry that 
can be measured with a transversely polarized target. 
 In the 
complete SIDIS cross section expression, 8 structure functions depending on the transverse target polarization
are present, yielding   8 asymmetries with   independent azimuthal modulations.
The Sivers asymmetries is the most famous one; it is related to 
the Sivers PDF, that gives a measurement of the correlation of the intrinsic transverse momentum of an unpolarized
quark in a transversely polarized nucleon.   
Results for the Sivers asymmetries have been provided for unidentified~\cite{Alexakhin:2005iw, Ageev:2006da} 
  and identified hadrons~\cite{:2008dn}.
The measured Sivers asymmetries are very small and compatible with zero, 
indicating also in this case a cancellation between the u and d-quark contribution in 
an isoscalar target.
   
\section{Hadron pair asymmetries}

Another mechanism sensitive to the transversity PDF is the 
two hadrons   production. 
The event selection for this analysis is the 
same as described for the single hadron analysis, 
apart for the lower cut on the hadron $z$, that has been released to 0.1.
All  the events with at least two hadrons in the 
final state have been considered, and a cut on the sum of the relative energies, $z=z_1+z_2<0.9$, 
has been applied to remove exclusive production.

As for the Collins mechanism, 
the fragmentation of a transversely polarized quark into a pair of hadrons 
is expected to depend on an azimuthal angle~\cite{Collins:1993kq, Artru:1995zu, Radici:2001na, Bacchetta:2003vn},  
yielding for the number of events:
 \begin{equation}
N = N_0 \cdot (1+f  \cdot P_t  \cdot D_{nn}\cdot A_{\phi_{RS}}  \cdot \sin(\phi_{RS})) 
\end{equation}
where $\phi_{RS}$ is defined as $\phi_{R}-\phi_{s'}$; $\phi_{R}$ is the azimuthal 
angle between the lepton scattering plane and the plane containing the virtual photon momentum \textbf{q} and the component $R_T$ of the relative hadron momentum \textbf{R}$=\frac{1}{2}$(\textbf{P$_1$}-\textbf{P$_2$}) which is perpendicular to the summed hadron momentum \textbf{P$_h$}$=$\textbf{P$_1$}+\textbf{P$_2$}. 
The asymmetry $A_{\phi_{RS}}$ is 
proportional to the transversity function convoluted with the fragmentation function describing the two hadrons production, $H_1^{\angle h}$:
\begin{equation}
A_{\phi_{RS}} \propto \frac{\sum_q e_q^2 \, \Delta_T q(x) \, H_1^{\angle h}(z, M_h^2)}       {\sum_q e_q^2 \,     q(x) \,       D_q^h(z, M_h^2)} 
\end{equation}
where $M_h^2$ is the invariant mass of the hadron pair. 

Two different analyses have been performed. 
 In the first one, the  hadron pair is  composed of
 a positive and  a  negative hadron: 
the hadrons have been identified with the RICH, so that four combinations have been considered: 
$\pi^+\pi^-$, $\pi^+ K^-$, $K^+ \pi^-$, $K^+ K^-$.
 In the second analysis, the hadrons have been ordered considering the 
leading and sub-leading hadrons (defined by the corresponding $z$); in this analysis, 
also combination of hadrons with the same charge are possible, leading to  16 combinations. 

All these  asymmetries have been measured as a function of $x$, $z$ and $M_h$.
No clear indication of a signal different from zero has been obtained from all the  measurements.

\section{Conclusions}
In 2002-2004 COMPASS has provided a complete set of measurement of transverse spin effects 
using a deuterium target. All the measured asymmetries 
are small and compatible with zero, implying a cancellation between the u and d-quark contribution in
a isoscalar target.

\begin{footnotesize}


\end{footnotesize}

\end{document}